\documentclass{appolb}
\usepackage{graphicx}

\begin{document}
\title{Collectivity in small and large amplitude microscopic mean-field dynamic %
\thanks{Presented at the XXII Nuclear Physics Workshop, Kazimierz, 2015, Poland}%
}
\author{Denis Lacroix, Yusuke Tanimura
\address{Institut de Physique Nucl\'eaire, IN2P3-CNRS, Universit\'e Paris-Sud, F-91406 Orsay Cedex, France}
\\
{Guillaume Scamps}
\address{Department of Physics, Tohoku University, Sendai 980-8578, Japan}
}
\maketitle
\begin{abstract}
The time-dependent energy density functional with pairing allows to describe a large variety of phenomena from 
small to large amplitude collective motion. Here, we briefly summarize the recent progresses made in the field using 
the TD-BCS approach. A focus is made on the mapping of the microscopic mean-field dynamic to the macroscopic 
dynamic in collective space. A method is developed to extract the collective mass parameter from TD-EDF. 
Illustration is made on the fission of $^{258}$Fm. The collective mass and collective momentum associated 
to quadrupole deformation 
 including non-adiabatic effects is estimated 
along the TD-EDF path. With these information, the onset of dissipation during fission is discussed.  
\end{abstract}
\PACS{PACS numbers come here}
  
\section{Introduction}
Recently, an effort has been made to give a unified description of small and large amplitude collective motion 
within the microscopic time-dependent energy density functional (TD-EDF) theory. The inclusion of pairing effects
in nuclear dynamic has opened new perspectives for the description of giant resonances \cite{Sca13b,Sca14}  or direct reactions like nucleon transfer \cite{Sca13a}.
Quite recently, the possibility to get physical insight on the fission process using microscopic transport models has been 
revisited  including pairing \cite{Sca15,Tan15} or not \cite{Sim14,God14}.

In the present article, some aspects of fission are discussed using the recently developed TD-BCS model. 
 \begin{figure}[htb]
  \begin{center}
\includegraphics[width=10cm]{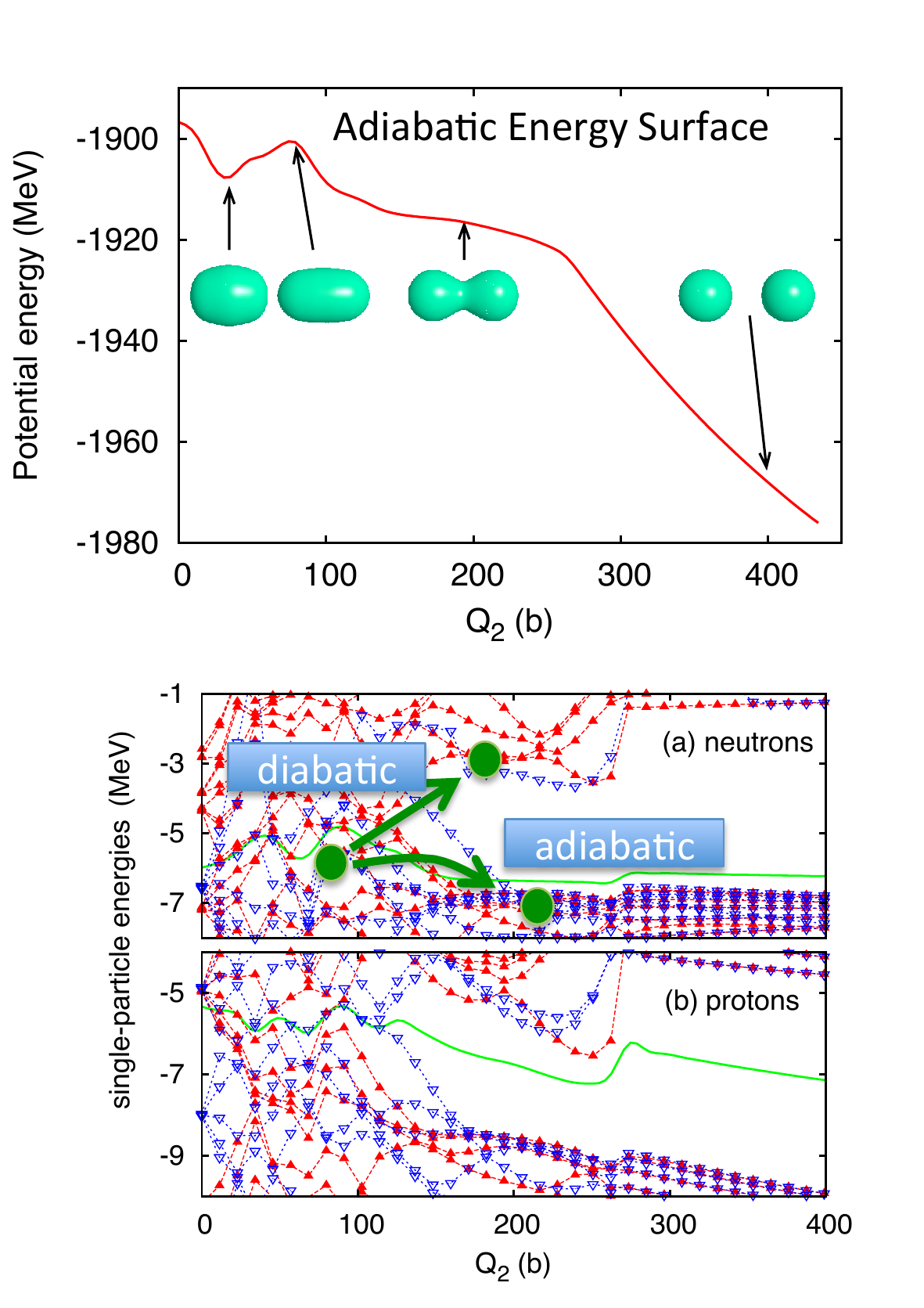} 
\end{center}
\caption{ (color online) Top: Adiabatic potential energy curve obtained for the case of compact symmetric fission 
in  $^{258}$Fm . The corresponding neutron and proton single-particle energies are shown in the bottom part.
See Ref.\cite{Sca15,Tan15} for details. The two extreme cases of adiabatic and diabatic motions are 
illustrated close to level crossing.}
\label{fig:pes}
\end{figure} 

\section{Fission dynamic with TD-BCS}

The TD-BCS approach is a microscopic method that solve the evolution of quasi-particle states 
in their self-consistent mean-field and pairing field. Technical aspects related to the TD-BCS we are using
are extensively described in Refs \cite{Sca13a, Eba10, Sca12}.
Some great advantages of this approach is that (i) it allows to treat 
nuclear dynamics from small to large amplitude collective motion (ii) it does not pre-select a priori 
specific collective degrees of freedom (DOFs). As a matter of fact, any collective DOF can play an important role 
as soon as it can acquire non-zero value consistently with the symmetries of the initial condition. (iii) it does not presuppose 
that the collective motion is adiabatic or not.  In the context of fission, it is still quite useful to  first consider 
the adiabatic energy landscape as a function of some collective DOFs, like elongation, multipole moments, ... 

\begin{figure}[htb]
  \begin{center}
\includegraphics[width=10cm]{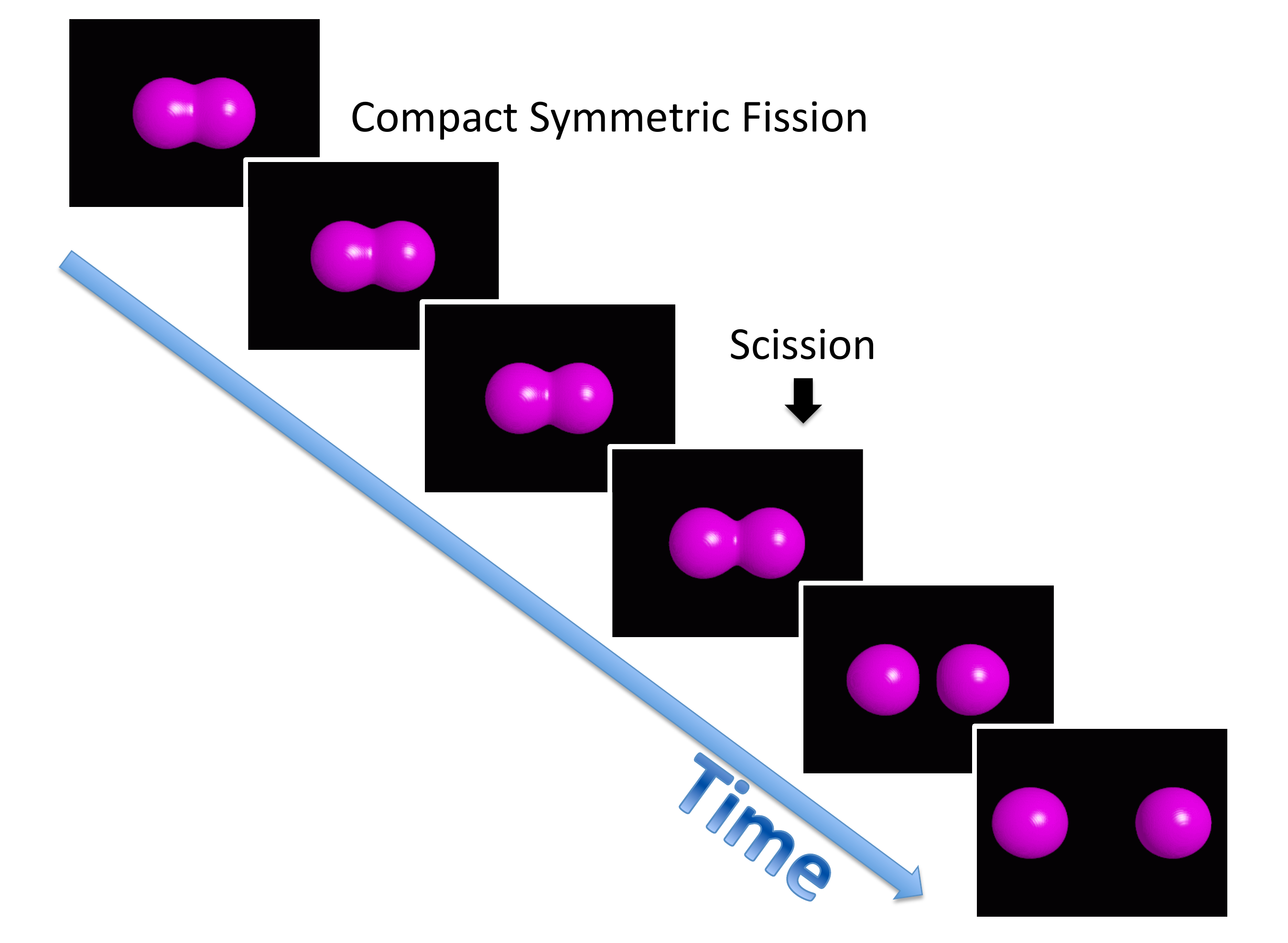} 
\end{center}
\caption{ (color online) Illustration of the density profiles obtained with TD-EDF at different time of the fission process.}
\label{fig:movie}
\end{figure}

An illustration is given in Fig. \ref{fig:pes} for the case of $^{258}$Fm fission. This curve is obtained here using the 
static version of TD-EDF with various constraint on the quadrupole moment.  
In the limit of very slow fission, it is expected that the dynamics directly reflects the motion along the adiabatic path.
However, starting from one of the point in the curve, there is no reason that the TD-EDF evolution follows this energy landscape.
Indeed,  the motion can eventually be rather fast especially close to the scission point where the slope of the energy
landscape change abruptly. The motion can be more complex than a collective motion in one dimension 
due to the possibility to excite other DOFs. The departure from adiabatic motion can directly be observed in the evolution 
of the nucleus density in TD-EDF (see Fig. \ref{fig:movie}).     
We see in this figure, that the two nuclei after scission can exhibit large octupole deformation. Such deformations 
are not observed along the adiabatic path.  

\section{Collective aspects of mean-field dynamic}

To further study some aspects of fission with TD-EDF and make connection with macroscopic models 
it is highly desirable to be able to define collective masses and momenta associated with a set of degrees 
of freedom. A method has been proposed in Ref. \cite{Tan15}.
\begin{figure}[htb]
 \begin{center}
\includegraphics[width=10cm]{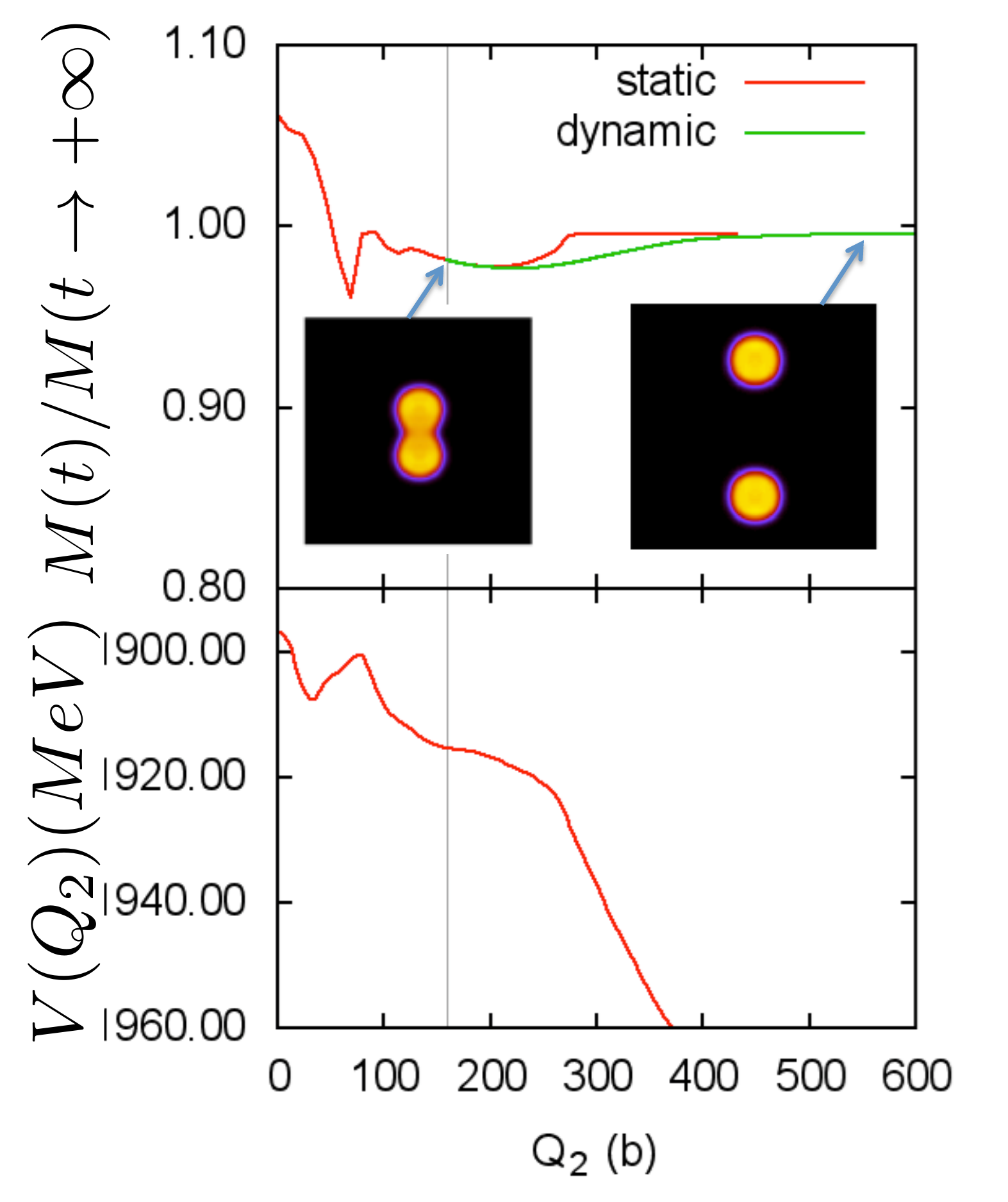} 
\end{center}
\caption{ (color online) Top: Evolution of the collective quadrupole mass along the adiabatic path (red) and along the 
TD-EDF path (green). Bottom: the corresponding position in the PES is recalled. }
\label{fig:mass}
\end{figure}
In general, the collective variable under interest $\hat Q$ is known explicitly (multipole moments, relative distance, mass asymmetry, neck, ...). 
Less easy is the knowledge of the associated collective momentum $\hat P$ and associated mass $M$. Several methods 
based on adiabatic approximation are usually used to construct these quantities. Here, we directly deduce them from TD-EDF evolutions.   

Assuming that the collective variable is local, and that the conjugated variables should fulfill the two conditions:
\begin{eqnarray}
\langle   [\hat Q_\alpha, \hat P_\alpha ] \rangle  &=&  i \hbar 
\end{eqnarray} 
and
\begin{eqnarray}
\frac{d \langle \hat Q \rangle}{dt}  &=&   -\frac{i}{2\hbar m} {\rm Tr} \left( \left[ Q , p^2\right]
\rho (t)  \right) \equiv  \frac{\langle \hat P\rangle}{M}
\end{eqnarray}
It was possible to prove that the collective momentum can be written as
\begin{eqnarray}
P  &\equiv&  - i \hbar \frac{M}{m} \left( \frac{\nabla^2 Q }{2} +\nabla Q . \nabla \right).
\end{eqnarray}
Here, the collective mass $M$ is given by:
\begin{eqnarray}
\frac{1}{M (t)} = \frac{1}{m} 
{\rm Tr} \left[ \rho(t) {\nabla Q} . {\nabla Q} \right]  .
\end{eqnarray}
This expression of the mass is sometimes also obtained in other approaches that 
specifically assume  adiabatic or diabatic motion. The great difference stems from 
the fact that here the density $\rho(t)$ entering in it is the TD-EDF one. Therefore, it can 
contain non-adiabatic effects as well as possible influence of other collective and non-collective 
DOFs. An illustration of the quadrupole collective mass obtained with the TD-EDF method is shown in Fig. 
\ref{fig:mass}. For comparison, the mass obtained assuming that the density identifies with the density along the 
adiabatic path is shown. While at initial instants, the system follows the adiabatic limit, when it approaches 
the scission, the collective motion accelerate and clear deviation from adiabaticity is observed.
 \begin{figure}[htb]
 \begin{center}
\includegraphics[width=10cm]{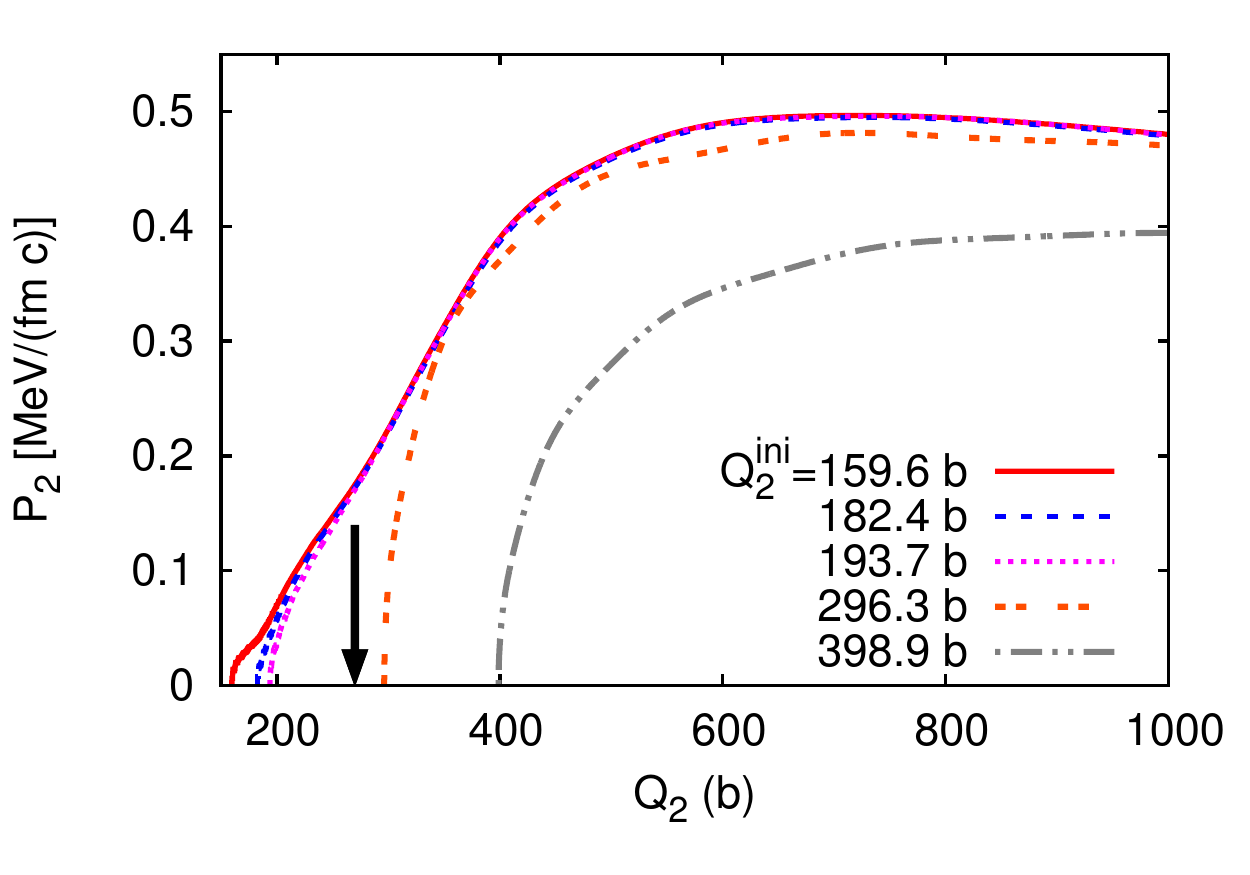} \\
\includegraphics[width=10cm]{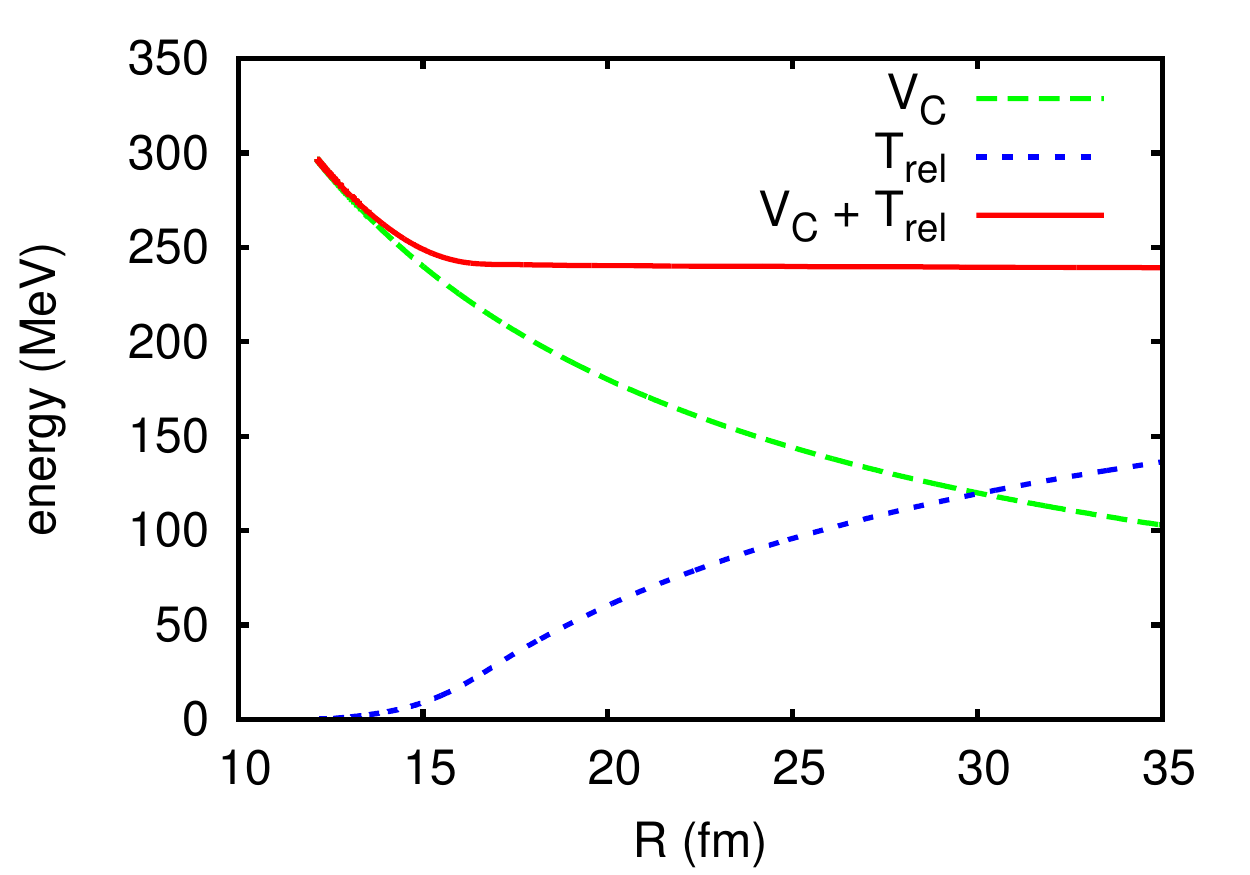}
\end{center}
\caption{ (color online) Top: Evolution of the collective quadrupole momentum as a function of the $Q_2$ 
for different initial conditions. Bottom: Energy balance between the kinetic energy associated to the relative 
motion of fissioning fragments and Coulomb energy. The sum of the two quantities is shown to saturate to the final total kinetic energy 
after scission point.  }
\label{fig:diss}
\end{figure}

\section{Dissipated energy along the fission path}

The possibility to access collective momentum is useful to get information on the energy 
balance during fission. In particular, it can give access to some dissipative aspects. 
We show in Fig. \ref{fig:diss} examples of evolution of the quadrupole momentum for different 
initial conditions taken for initial values of $Q_2$ lower or greater than the scission point $Q^{\rm sc}_2$. 
The most important feature  is that, for $Q_2 \le Q^{\rm sc}_2$, all curves seem to become identical before 
reaching scission. This could only be understood  assuming that the system is strongly damped 
at the early instant of its dynamical evolution and rapidly end up along the same dynamical path. 
It is worth mentioning that this path does not necessarily match with the one displayed in Fig.
\ref{fig:pes}. 

Starting from this finding, and from the knowledge of the collective momentum. 
It was possible to extract the total energy dissipated along the fission path \cite{Tan15}. It was shown that 
this dissipated energy is quite large and can approach 10 \% of the final total kinetic energy 
TKE of the daughter nuclei after fission. The TKE obtained in the symmetric compact fission of 
$^{258}$Fm is shown in bottom panel of Fig. \ref{fig:pes} and is close to 200 MeV. Therefore, around 
 20 MeV of the initial energy is dissipated during the fission. In the present calculation, this 
 energy is transferred to other internal DOFs of the two fissioning nuclei and can eventually lead 
 to particle evaporation.
 
\section{Summary}

In the present article, we have illustrated the fission of a superfluid nucleus 
using the TD-BCS approach. A method is used to get macroscopic information, like collective momentum and mass 
from the microscopic evolution. In particular, it is shown that the collective mass deviates 
from the adiabatic limit, especially close to the scission point where the evolution 
is faster. The sharing of the initial energy between total kinetic energy of final fragments and 
internal excitations is also studied. It is seen that almost 10 \% of the TKE was dissipated. This
dissipation occurs at the very first instants of the dynamical evolution.

\section{ACKNOWLEDGMENTS}
We  thank P2IO
for Y. T. postdoctoral grant. G.S. acknowledges the Japan Society for the Promotion of Science for the 
JSPS postdoctoral fellowship for foreign researchers. This work was supported by Grant-in-Aid for JSPS Fellows No. 14F04769.

\end{document}